\documentclass[a4paper,12pt]{article}
\usepackage{epsfig} \usepackage{amssymb} \usepackage{amsfonts}
\def\eV{{\rm e\kern-0.12em V}} \def\GeV{{\rm G}\eV} \def\MeV{{\rm M}\eV}
\def\MSbar{\relax\ifmmode\overline{\rm MS}\else{$\overline{\rm MS}${ }}\fi}
\def\msbar{\relax\ifmmode\overline{\rm MS}\else{$\overline{\rm MS}${ }}\fi}
\def\alphan{\relax\ifmmode{\alpha_{\rm an}}\else{$\alpha_{\rm an}${ }}\fi}
\def\alphas{\relax\ifmmode{\alpha_{\rm s}}\else{$\alpha_{\rm s}${ }}\fi}
\def\asmz{\relax\ifmmode\bar \alpha_s(M_Z^2)\else{$\bar \alpha_s(M_Z^2)${ }}\fi}
\def\albars{\relax\ifmmode{\bar{\alpha}_s}\else{$\bar{\alpha}_s${ }}\fi}
\def\tildal{\relax\ifmmode{\tilde{\alpha}}\else{$\tilde{\alpha}${ }}\fi}
\def\tildals{\relax\ifmmode{\tilde{\alpha}(s)}\else{$\tilde{\alpha}(s)${ }}\fi}
\def\asQ{\relax\ifmmode\bar{\alpha}_s(Q^2)\else{$\bar{\alpha}_s(Q^2)${ }}\fi}
\def\agoth{\relax\ifmmode{\mathfrak A}\else{${\mathfrak A}${ }}\fi}
\def\pisq{\relax\ifmmode{\pi^2}\else{${\pi^2}${ }}\fi}
\def\agothk{\relax\ifmmode{\mathfrak A}_k\else{${\mathfrak A}_k${ }}\fi}
\def\acal{\relax\ifmmode{\cal A}\else{${\cal A}${ }}\fi}
\def\acalk{\relax\ifmmode{\cal A}_k\else{${\cal A}_k${ }}\fi}
\newcommand{\beq}{\begin{equation}} \newcommand{\eeq}{\end{equation}}
\newcommand{\beglab}{\begin{equation}\label}
\begin{document}
\begin{center}
{\large\sf Analytic Perturbation Theory for QCD observables \\}
\bigskip

D.V. Shirkov \\
\medskip

{\it Bogoliubov Laboratory, JINR, 141980 Dubna, Russia\\
e-address: shirkovd@thsun1.jinr.ru}
\end{center} \smallskip

\centerline{\bf Abstract}\smallskip

 {\small
     The  connection
between ghost--free formulations of RG--invariant perturbation
theory in the both Euclidean and Minkowskian regions is
investigated. Our basic tool is the ``double spectral
representation", similar to the definition of Adler function, that
stems from first principles of local QFT. It relates real
functions defined in the Euclidean and Minkowskian regions.  \par

 On this base we establish a simple relation between \par
--- The trick of resummation of the $\pi^2$--terms (known from early 80s) for
the invariant QCD coupling and observables in the time-like region and \par
---  Invariant Analytic Approach (devised a few years ago) with the ``analyticized"
coupling $\alpha_{\rm an}(Q^2)$ and nonpower perturbative expansion for
observables in the space-like domain which are free of unphysical singularities . \par

  As a result, we formulate a self--consistent scheme --- Analytic
Perturbation Theory (APT) --- that relates a renorm--invariant, effective
coupling functions $\alphan(Q^2)\,$ and $\tildal(s)\,,$ as well as non--power
perturbation expansions for observables in both space-- and time--like domains,
that are free of extra singularities and obey better convergence in the
infrared region. \par

  Then we consider the issue of the heavy quark thresholds and devise a
global APT scheme for the data analysis in the whole accessible space-like
and time-like domain with various numbers of active quarks. \par

  Preliminary estimates indicate that this global scheme produces results a
bit different, sometimes even in the five-flavour region, on a few per cent
level for $\bar{\alpha}_s$ -- from the usual one, thus influencing the
total picture of the QCD parameter correlation.} \newpage

\section{Introduction \label{s1}}
\subsection{Preamble: perturbation theory and \albars \label{ss1.1}}

 The issue of the strong interaction behavior at the low and medium energy
 $W=\sqrt{s}\,$  and momentum transfer $Q=\sqrt{Q^2}\,$ attracts more
and more interest along with the further experimental data accumulation.

As a dominant means of theoretical analysis, here one uses the perturbative
QCD (pQCD), in spite of the fact that in the given domain the power
expansion parameter \albars is not a ``small enough" quantity. Physically,
this region corresponds to three $(f=3)\,$ and four $(f=4)\,$ flavors
(active quarks). Just in the three--flavour region there lie unphysical singularities
of central theoretical object --- invariant effective coupling \albars.
These singularities, associated with the scale parameter $\,\Lambda_{f=3}
\simeq 350\,$ MeV, complicate theoretical interpretation of data in
the ``small energy" and ``small momentum transfer" regions $(\,\sqrt{s},\,
Q\equiv\sqrt{Q^2} \lesssim 3\Lambda_3\,.)$ On the other hand, their
existence contradicts some general statements of the local QFT. \par

 It is important to notice that in the current literature for the
effective QCD coupling in the time--like domain $\alphas(s)\,;~s=W^2\,$
one uses literally the same expression, like one in the Euclidean domain.
By the way, implanting the mentioned singularities into the three--flavor
region of small energies $\,W\simeq 350\,$ \MeV. \par

  Meanwhile, the notion of invariant electron charge (squared)
$\,\bar{\alpha}(Q^2)=\bar{e}^2(Q)\,$ in QED has initially been defined in
the early papers\cite{rg56} on renormalization group (RG) only in the
space-like region in terms of a product of real constants $\,z_i\,$ of finite
Dyson renormalization transformation. Just the Euclidean invariant charge
$\,\bar{e}(Q)\,$ is related by the Fourier transformation to the space
distribution $\,\bar{e}(r)\,$ of the electric charge (around a point
``bare" electron) introduced by Dirac\cite{dirac34}. \par

  Analogous motivation in the RG formalism (for detail, see chapter
``Renormalization group" in the text\cite{kniga}) underlies a more
general notion of invariant coupling $\,\bar{g}(Q)\,,$ {\it defined only
in the space--like domain.} Inside the RG formalism, there is no simple
means for defining  $\,\bar{g}\,$ in the time--like region. \par

  Nevertheless, in modern practice, inspired by ``highly authoritative
reviews"\cite{pdg00,siggi00} and some monographs (like \cite{bard99})
one uses the same singular expression for the QCD effective coupling
\albars both in the space-- and time--like domains. \par                        % 113

    Technically, this ``implanting" of the \albars Euclidean functional form
into Minkowskian is accompanied by some modification of numerical expansion
coefficients. To the initial coefficient calculated by Feynman diagrams,
one adds specific terms (containing $\pi^2\,$ and its powers) with
coefficients of some lower orders. These ``$\pi^2\,$--terms" are the only
``atonement for the Styx river crossing" from the Euclid realm to the
Minkowski domain.\par \medskip

\subsection{Time--like region,  $\pi^2\,$--terms}
 Meanwhile, as it easy to show, the ``$\pi^2$--procedure" is valid only at
small parameter $\pi^2/\ln^2(s/\Lambda^2)\,$ values, that is in the region
of high enough energies $W\gg\Lambda e^{\pi/2}\simeq 3\,\GeV\,$. \par

Here, it is useful to restore the ideas proposed by Radyushkin\cite{rad82}
and Krasnikov---Pivova\-rov\cite{kras82} (RKP procedure) at the beginning
of the 80s.

 To introduce an invariant \albars in the time--like region, both the
authors used integral transformation $\,{\bf R}$, ``reverse" to the
Adler function definition. The last one can be treated as the definition
of integral operation
\begin{equation}\label{d-trans}
R(s) \to D(z) = Q^2\int^{\infty}_0 \frac{d s}{(s+z)^2}\,R(s)\,                % 1
\equiv {\cal\bf D} \left\{ R(s)\right\}\,, \eeq
transforming a real function $\,R(s)\,$ of positive (time--like)
argument into the function $\,D(z)\,$ defined in the cut complex plane
with analytic properties adequate to the K\"all\'en--Lehmann
representation. In particular, $\,D(Q^2)\,$ is real at the positive
semiaxis $z=Q^2+i0\,; Q^2\geq 0\,.$\par

The reverse operation ${\cal\bf R}\,$ can be expressed via the contour integral
$$
R(s)=\frac{i}{2\pi}\,\int^{s+i\varepsilon}_{s-i\varepsilon}\frac{d
z}{z}\, D_{\rm pt}(-z)\equiv{\bf R}\left[D_{\rm }(Q^2)\right]\,.$$

By operation ${\cal\bf R}\,$ one can define RG--invariant, effective
coupling $\tildal(s)={\bf R}\left[\albars(Q^2)\right]\,$ in the
time--like region. %\par
   A few simple examples are in order     \smallskip

---  For the one--loop effective coupling $\albars^{(1)}=\left[\beta_0
\ln(Q^2/\Lambda^2)\right]^{-1}\,$ one has \cite{schr2,rad82,js95}\footnote{
Later on, this idea has been discussed by several known authors ---
see Refs. \cite{bj89} --- \cite{ms97}.}
\beq \label{tildal1}
{\bf R}\left[\albars^{(1)}\right]\,=\tildal^{(1)}(s)=\frac{1}{\beta_0}
\left[\frac{1}{2}-\frac{1}{\pi}\arctan\frac{L}{\pi}\right]_{L>0}=\frac{1}
{\beta_0\pi}\arctan\frac{\pi}{L}\,;\quad L=\ln\frac{s}{\Lambda^2}\,.\eeq

--- At the two--loop case, to the popular approximation
$$\beta_0\bar{\alpha}_{s,pop}^{(2)}(Q^2)=\frac{1}{l}-b_1(f)
\frac{\ln l}{l^2}\:; \quad l=\ln\frac{Q^2}{\Lambda^2}\,$$
there corresponds \cite{rad82,brs00}
\beq\label{tildal2pop}
\tildal^{(2)}_{pop}(s)=\left(1+\frac{b_1 L}{L^2+\pi^2}
\right)\tildal^{(1)}(s) -\frac{b_1}{\beta_0}\frac{\ln\left[
\sqrt{L^2+\pi^2}\right]+1}{L^2+\pi^2}\,. \eeq                             %  3

 Both the expressions (\ref{tildal1}) and (\ref{tildal2pop}) are monotonously
decreasing with finite IR limit $\,\tildal(0)=1/\beta_0(f=3)\simeq1.4\,.$

--- At the same time, square and cube of $\albars^{(1)}\,$ transform
into simple ``pipizated" expressions\cite{rad82,kras82}
\beq\label{agoth23-1}
\agoth_2^{(1)}(s)\equiv {\bf R}\left[\left(\albars^{(1)}\right)^2\right]
=\frac{1}{\beta_0^2\left[L^2+\pi^2\right]}\,\quad \mbox{¨} \quad
\agoth_3^{(1)}(s)=\frac{L}{\beta_0^3\left[L^2+\pi^2\right]^2}\,,\eeq
which {\it are not powers} of  $\tildal^{(1)}(s)\,.$ \par

\begin{quote}
{\small\it Note also that transition from singular \albars and its powers to
``pipizated" expressions, that is operation ${\bf R}$, can be performed
\cite{bj89,dmw96} by the differential operator
$${\cal R}= \frac{\sin \pi {\cal P}}{\pi {\cal P}}\,; \quad {\cal P}=
Q^2\,\frac{d}{d Q^2}\:\quad \mbox{with the substitution}\quad Q^2\to s\,,
\ \ \mbox{e.g.,} \ \  {\cal R}\asQ = \tildal(s)\,.$$ }
\end{quote}\par \smallskip

   The most remarkable feature of all presented expressions for \tildals
and $\agoth_{k}(s)$ (valid in a more general case) is the absence of
unphysical singularity (the ``log pole" at the one--loop case)\footnote{
 This feature was not mentioned in the pioneer papers of the 80s we have
cited above.} which is  ``screened" by $\pi^2\,$--contributions.\par
  Besides,  a common ``Euclidean" perturbation expansion
\beq\label{standQ}
 D_{\rm pt}(Q^2)=1+ \sum_{k\geq 1}^{} d_k\,\albars^k(Q^2) \eeq
in powers of the standard RG--summed effective coupling $\asQ,$ with its
unphysical singularities in the IR region (at $Q^2\leq\Lambda^2_3\,,$)
being transformed by ${\bf R}$ to the time--like region, transits  into
the asymptotic expansion over a nonpower set of functions
\begin{equation}\label{Rpi}
R_{\pi}(s)\equiv {\bf R}\left[D_{\rm pt}(Q^2)\right]=1+
\sum_{k\geq 1}d_k{\mathfrak{A}}_k(s)\,;\quad
~\mathfrak{A}_k(s)={\bf R}\left[\albars^k(Q^2)\right]\,,\eeq
with better properties of decreasing\cite{rad82} of subsequent terms .\par

 At the same time, higher functions, like (\ref{agoth23-1}),
vanish $\agothk(0)=0\,; \ k \geq 1\,$ in the IR limit. \par

 On the other hand, in the UV region at $\ln(s/\Lambda^2)\gg \pi\,,$ i.e.,
for $W\gg \Lambda e^{\pi/2}\simeq 3\, \GeV\,,$ the functions \tildal and
$\agoth_k$ can be represented as a series in powers of the parameter $\pi^2/L^2,
~ L=\ln(s/\Lambda^2)\,.$ Such expressions sometimes can be reformulated
into expansions in powers of \alphas. For instance, in the one-loop case
\beglab{tildal-exp}
 \tildal^{(1)}(s) \simeq \frac{1}{\beta_0 L} - \frac{\pi^2}{3 L^2}+       %eq 7
 \frac{\pi^4}{5 L^4}+ \simeq \albars^{(1)}(s)-
\frac{\pi^2\beta_0^2}{3}\,\left(\albars^{(1)}(s)\right)^3+
\frac{\pi^4\beta_0^4}{5}\,\left(\albars^{(1)}(s)\right)^5
\eeq \smallskip

 Without going into detail, note\footnote{See, also the first version of this
paper \cite{tow00}. A more minute numerical information on functions
\tildal, \alphan, $\agoth_{2,3}$ and $\acal_{2,3}$ can be found in recent
paper by Magradze \cite{mag00}.} that, qualitatively, the functions \agothk
behave very similarly to the functions \acalk involved into non-power
Euclidean asymptotic expansion for observables\cite{lmp99tmp} arising in
the Analytic Perturbation Theory (APT) -- see below Section
\ref{ss1.3} and Figure 2. In particular, they oscillate at small argument
values and form an asymptotic set \`a l\'a Erd\'elyi. \par

 As it follows from eqs. (\ref{tildal1}) and (\ref{agoth23-1}), one--loop
``pipizated" functions satisfy the recursion relation  $\,(d/d L)
\mathfrak{A}^{(1)}_k(s)=-k\,\beta_0\,\mathfrak{A}^{(1)}_{k+1}(s)\,$
which is analogous to the one-loop differential equation for the invariant
coupling. According to \cite{lmp99tmp}, this recursion is valid for
analyticized functions  $\acalk^{(1)}$. \par

  Quite recently \cite{mag00} the two--loop generalization has been found % p 5
\beq\label{rec2}\frac{1}{k}\frac{d}{d L}\mathfrak{A}^{(2)}_k(s)=
-\,\beta_0\,\mathfrak{A}^{(2)}_{k+1}(s)\,
-\,\beta_1\,\mathfrak{A}^{(2)}_{k+2}(s)\,. \eeq                       % 8
 It can be considered as (at $\,k=1\,$) a mould of two--loop
differential equation for \albars. Analogous relation is valid for
analyticized \acalk. \medskip

  For the reverse transition from Minkowski to Euclid, one could try to use
the transformation $\,{\cal\bf D}\,$ defined by (\ref{d-trans}). However,
it is evident that we shall not return to the initial coupling \albars and
to series in its powers (\ref{standQ}). To elucidate the issue, it is useful
to turn to the foundation of the Invariant Analytic Approach mentioned above.

\subsection{Space--like region : APT  \label{ss1.3}}                  % 251

  Indeed, as it has been well-known from the late 50s \cite{bls59}, there
exists a method
of getting rid of Euclidean unphysical singularities by combining RG-summed
expressions with K\"all\'en--Lehmann analytical representation for \asQ in
the $\,Q^2\,$ variable. In the mid-90s this idea was used in
QCD~\cite{rapid96,prl97,ss98pl} under the name of Invariant Analytic
Approach. Its further development and application to perturbative expansion
for observables yielded Analytic Perturbation Theory --- \cite{npb98}.\par

  We remind here the basic features and results of APT ( --- see also a
recent review \cite{ss99tmp}). \smallskip

 By combining three elements \par
{\bf 1.} Usual Feynman perturbation theory for effective coupling(s) and
observables,\par
{\bf 2.} Renormalizability, i.e., renormalization--group (RG) invariance,
and \par
{\bf 3.} General principles of local QFT --- like causality, unitarity,
Poincar\'e invariance \par and spectrality --- in the form of spectral
representations of the K\"allen--Lehmann and \par Jost--Lehmann--Dyson type
\par \smallskip

it turns out to be possible to formulate an {\it Invariant Analytic Approach}
(IAA) for the pQCD invariant coupling and observables in which the central
theoretical object is a {\it spectral density}.\par \smallskip

\begin{itemize}
\item Being calculated by the usual RG--improved perturbation theory, it
defines and relates $\,Q^2$--analytic,
RG-invariant expressions for effective RG--invariant coupling and
perturbative observables in the Euclidean channel. \par
\smallskip

\item In particular, the IAA results in the modified ghost-free expression
for the invariant QCD coupling  $\,\alpha_{\rm an}(Q^2;f)\,$ which is free
of ghost troubles and obey reduced~\cite{prl97} -- \cite{mss99gls}
higher--loops and renormalization--scheme sensitivity\footnote{This
analyticized QCD coupling \alphan has been successively
used\cite{nicos99,nicos00} in analysis of the pion and
$\gamma^*\gamma \to \pi^o$ formfactors.}. See, Fig.1.  \par

\item
 The IAA change the structure of perturbation expansion for observables:
Instead of common power series, as a result of integral transformation,
there appears non--power asymptotic expansion~\cite{lmp99tmp} \'a la
Erd\'elyi over the sets of specific functions  ${\cal A}_k(Q^2;f)\,,$ free
of unphysical ghosts. These functions are defined via integral
transformations of related powers $\alpha_s^k(Q^2;f)\,$ in terms of
relevant spectral densities. This nonpower expansion for an observable,
with the coefficients extracted from the relevant Feynman diagrams,
we call the {\sf \/Analytic Perturbation Theory.} \par

 At small and moderate arguments, \/\acalk diminish with the $\,k\,$ growth
much quicker than the powers of $\,\alpha_{\rm an}^k\,$ (and even oscillate
in the region $ \sqrt{s}, Q\simeq \Lambda\,,$) thus improving essentially
the convergence of perturbation expansion for observables. \par
\end{itemize} \smallskip

  The first purpose of this work is to elucidate relation between the
Radyushkin--Krasnikov--Pivovarov procedure leading to effective summation of
$\pi^2$--terms (``pipization" trick)\cite{rad82,kras82} for observables and
the Solovtsov\cite{js95,ms97} construction of the effective QCD coupling
within the IAA scheme in the $s$--channel. \par

 In the course of this analysis --- see Section \ref{s2} --- we discuss
the APT proliferation to the time--like region, remind a spectacular
effect of ``distorting mirror" correlation\cite{mo98} between analyticized and
pipizated invariant QCD couplings in space-like $\,\alpha_{\rm an}(Q^2;f)\,$
and time-like $\,\tildal(s;f)\,$ regions (see Fig.1 below), and establish this
effect for corresponding expansion functions ${\cal A}_k(Q^2;f)$ and
${\mathfrak A}_k (s;f)$ -- see Fig.2. \par \smallskip

  Then, in Section \ref{s3}, we consider an the transition across the
heavy quark thresholds, to construct a ``global" picture for the whole
physical region $M_{\tau}\lesssim\sqrt{s}, Q\lesssim M_Z$ --- see Fig.2.\par

It should be noted, that all precedent papers Refs.\cite{kras82} --
\cite{nesholom} dealt only with the massless quarks  in the case with fixed
flavour number $\,f\,$. This can be justified, to some extent, when analyzing
inside a narrow interval of the relevant energy $\sqrt{s}\,$ or momentum
transfer $Q\,$ values. Meanwhile, the ultimate goal of all the pQCD is a
correlation of effective coupling values extracted from different experiments.
 \bigskip

 Main results of this investigation are reviewed in the Conclusion.

\section{Self-consistent scheme for observables\label{s2}}
\subsection{Modification of the APT \label{ss2.1}}
 As it has been mentioned above, applying operation ${\cal\bf D}\,$ to
\tildal does not restore a usual effective coupling as far as representation
(\ref{d-trans}) is not compatible with ghost singularity of \albars.\par
  Instead, we arrive at
\begin{equation}\label{al-dipole}{\cal\bf D}\left\{\tildal(s;
f)\right\}=\frac{Q^2}{\pi}\int^{\infty}_0\frac{d s}
{(s+Q^2)^2}\, \tildal(s; f)\,\equiv\alphan(Q^2; f) \;,\eeq
i.e., to effective Euclidean coupling \alphan of APT. This simple fact has
first been established in \cite{js95}. We see that operations
$\,{\cal\bf D}\,$ and $\,{\bf R}\,$ relate ``pipizated" and ``analyticized"
coupling functions in space-- and time--like regions. Hence, in this case
$\,{\bf R}={\cal\bf D}^{-1}$. Note, however, that the relation
$\,{\cal\bf D}{\bf R}={\bf 1}\,$  is valid {\it only for the class of
functions} $F(Q^2) \in {\cal\bf C}_{KL}\,$ {\it satisfying the
K\"allen--Lehmann representation.}\par

  Now, we have the possibility of extending the APT to the time-like region.         % p 7
 We shall do it in the form of a recipe, using operation of analyticization
\beq\label{operA}                                                              %  10
F(Q^2)\,\to F_{\rm an}(Q^2)\,={\cal\bf A}\cdot F(Q^2)\,,\eeq
first introduced in Refs.\cite{rapid96,prl97} in terms of the            % p 7
 K\"allen--Lehmann representation
\beq\label{defFan} F_{\rm an}(Q^2)=\frac{1}{\pi}\int\limits_{0}^{\infty}
\frac{d\sigma}{\sigma+Q^2}\,\rho_{\rm pt}(\sigma)\,;\quad \rho_{\rm pt}
(\sigma)=\Im\:F(-\sigma) \,,\eeq
with spectral density defined via straightforward continuation
of $F$ on the cut. \smallskip

 Relations (\ref{operA}) and (\ref{defFan}) together define {\bf €}
the analyticization operation. Now, we can formulate the APT anew. \smallskip

  {\sf Firstly,} one has to transform the common singular coupling
function \asQ or some power expansion of an observable
\begin{equation} \label{Dpt}
D_{\rm pt}(Q^2)=1+ \sum_{k\geq 1} d_k\,\albars^k(Q^2;f) \eeq
into the corresponding analytic Euclidean expression \alphan or
$\,D_{\rm an}(Q^2)\,,$ free of ghosts
\begin{equation}\label{Dan}                                           % eq 13
D_{\rm an}(Q^2;f)=1+\sum_{k\geq 1} d_k\,{\cal A}_k(Q^2;f)\:;  \quad
\alpha_{\rm an}(Q^2;f)= {\cal A}_1 (Q^2;f)\,, \eeq
\begin{equation}\label{defAk}
{\cal A}_k(x;f)=\frac{1}{\pi}\int\limits_{0}^{\infty}\frac{d\sigma}
{\sigma+x}\:\rho_k(\sigma;f)\,;~~\rho_k(\sigma;f)
=\Im\left[\albars^k(-\sigma;f) \right]\, \end{equation}
with spectral densities $\rho\,,\rho_k $ introduced according to   % c 8
 (\ref{defFan}). \par \medskip

  {\sf Secondly,} by operation ${\bf R}\,$ one defines\cite{js95} in the
Minkowskian region invariant coupling function\footnote{As it follows from
this expression, the spectral function
can be considered as a beta--function. However, contrary to  Schwinger's
hope\cite{schwin}, this $\rho(s; f)\,,$ being a spectral function for the
Euclidean invariant coupling, happens to be the RG generator for another,
Minkowskian, invariant coupling\cite{ms97}.}
\beq\label{tildal_f}                                                    % 15
\alphan(Q^2;f)\to\tildal(s;f)={\bf R}\left[\alphan\right] =
\int\limits^{\infty}_{s}\frac{d\,\sigma}{\sigma}\rho(\sigma; f)\,\eeq
or some other quantity like
$$R_{\pi}(s)\equiv {\bf R}\left[D_{\rm pt}(Q^2)\right]=1+ \sum_{k\geq
1}d_k{\mathfrak{A}}_k(s)\,;\quad~\mathfrak{A}_k(s)={\bf R}
\left[\albars^k(Q^2)\right]\,\eqno(6) $$
with
\begin{equation}\label{m-s}                                         % 16
\mathfrak{A}_k(s)=\int\limits^{\infty}_{s}\frac{d \sigma}{\sigma}
\rho_k(\sigma)\,\,; ~~\rho_k(\sigma)
=\Im\left[\alpha_s^k(-\sigma)\right]\,. \eeq \smallskip

{\sf Finally,} we have a simple possibility of reconstructing an Euclidean
object from the corresponding Minkowskian one with the help of the dipole
operator $\,{\bf D}\,$ like
$$\,\alphan(Q^2; f)=  {\cal\bf D}\left\{\tildal(s;f)\right\}\;.$$
 In particular, by substituting $\tildal^{(1)}(s; f)$ into the integrand,
we obtain after integration by parts
 $$
{\cal\bf D}\left\{\tildal^{(1)}\right\}=\frac{Q^2}{\pi\beta_0}
\int^{\infty}_{0}\frac{d\sigma}{(\sigma+Q^2)^2}\cdot\left(\frac{1}
{2}-\frac{1}{\pi}\arctan\frac{\ln(\sigma/\Lambda^2)}{\pi}\right)=$$
\begin{equation}\label{an1Q}                                        % eq 17
=\frac{1}{\beta_0}\left[\frac{1}{\ln Q^2/\Lambda^2}\,+\,\frac{
\Lambda^2}{\Lambda^2-Q^2}\right]~=\alpha^{(1)}_{\rm an}(Q^2,f)\,. \eeq

 This simple calculation elucidates the relation between ghost--free
expressions in the Min\-kow\-skian and Euclidean regions. They are related
by a reverse transformation as well. For instance, in accordance with
(\ref{tildal_f}),
 $$ \tildal^{(1)}(s;f)={\bf R}\left[\alphan^{(1)}(Q^2;f)\right]\,. $$

In Fig.1, we give a concise summary of the IAA results for invariant
analytic couplings $\alphan(Q^2,3)$ and $\tildal(s,3)$ calculated for one--
, two-- and three--loop cases in both the Euclidean and Minkowskian domains. \par
%%%%%  FIGURE-1   ===  %%%%%%%%%%%%%%%%
 \begin{figure}[th]\label{fig1}
 \unitlength=1mm
   \begin{picture}(0,100)                                   %
   \put(10,1){%                                             %
   \epsfig{file=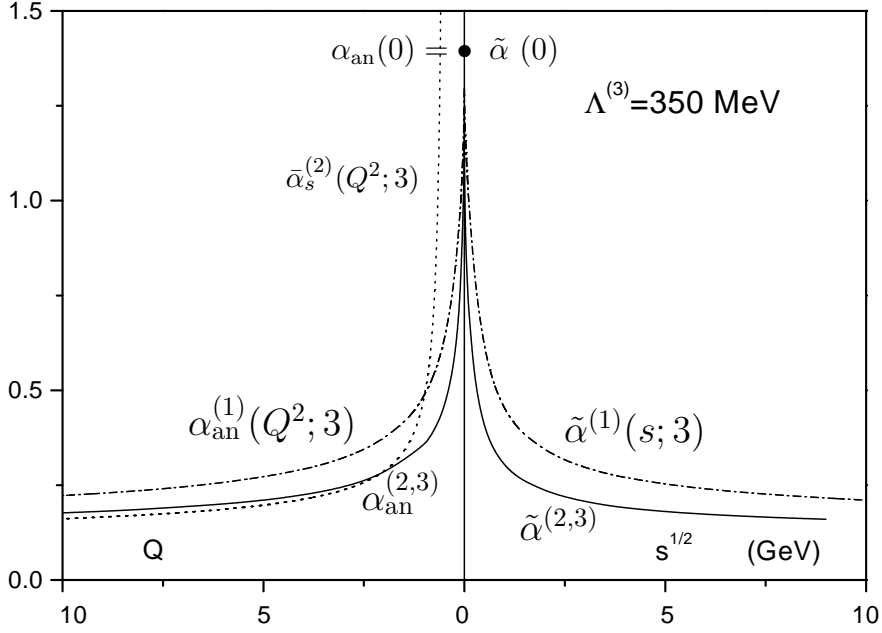,width=14cm}%                     %
   }                                                        %
  \put(82.7,87){$\bullet$}%                                 %
  \put(66,87){$\alphan(0)=$}%                               %
  \put(87,87){\tildal(0)}%                                  %
   \put(60,70){\small $\bar\alpha_s^{(2)}(Q^2;3)$}%         %
   \put(47,37){\large $\alpha_{\rm{an}}^{(1)}(Q^2;3)$}%     %
   \put(70,27){\large $\alpha_{\rm an}^{(2,3)}$}%           %
   \put(97,36){\large $\tilde{\alpha}^{(1)}(s;3)$}%         %
   \put(91,23){\large $\tildal^{(2,3)}$}%                   %
   \end{picture}                                            %
 \caption{\sl Space-like and time-like invariant analytic
 couplings in a few GeV domain} %
 \end{figure}
%%%%%%%%%%%%%%%%%%%%%%%%%%%%%%%%%%
 Here, the dash--dotted curves represent the one-loop IAA approximations
(\ref{tildal1}) and (\ref{an1Q}). The solid IAA curves are based on the
exact two-loop solutions of RG equations\footnote{ As it has recently been
established the exact solution to the two-loop RG differential equation for
the invariant coupling can be expressed in terms of a special function $W$,
the Lambert function, defined by the relation $ W(z) e^{W(z)}=z $ with an
infinite number of branches $\,W_n(z)$. For some details of analyticized
and pipizated solutions expressed in terms of the Lambert function, see
Refs. \cite{ggk98,badri98,badri99,mag00,kura00,Lambert}.} and approximate
three--loop solutions in the \msbar scheme. Their remarkable coincidence
(within the 1--2 per cent limit) demonstrates reduced sensitivity of the
IAA with respect to the higher--loops effects in the whole Euclidean and
Minkowskian regions from IR to UV limits. \par

 For comparison, by the dotted line we also give a usual \asQ two-loop
effective QCD coupling with a pole at $Q^2=\Lambda^2\,.$ \par

 As it has been shown in \cite{prl97,ss99tmp,MSS-PL97-tau}, relations
parallel to eqs.(\ref{tildal_f}) and (\ref{al-dipole}) are valid for
powers of the pQCD invariant coupling. This can be resumed in the
form of a self-consistent scheme. Consider now new functional sets
of nonpower perturbation expansions.

\subsection{Expansion of observables over nonpower sets
$\left\{{\cal A}\right\}$ and $\left\{{\mathfrak A}\right\}$\label{ss2.2}}         % 479

 To realize the effect of transition from expansion over the ``traditional"
power set $$\,\left\{\albars^k(Q^2,f)\right\}
\,=\,\asQ,\,\albars^2,\, \dots \albars^k \dots \, $$ to expansions
over nonpower sets in the space-like and time-like domains $$
\left\{{\cal A}_k(Q^2,f)\right\}\,=\,\alpha_{\rm an}(Q^2,f),\, {\cal
A}_2(Q^2,f) ,\,{\cal A}_3 \dots~;~\left\{{\mathfrak A}_k(s,f) \right\}
=\,\tilde{\alpha}(s,f),\,{\mathfrak A}_2(s,f),\,{\mathfrak A}_3\dots
\,,$$ it is instructive to learn properties of the latter.

  In a sense, both nonpower sets are similar \par
--- They consist of functions that are free of unphysical singularities. \par
--- First functions, the new effective couplings, ${\cal A}_1 =
\alpha_{\rm an}$ and ${\;\mathfrak A}_1 = \tilde{\alpha}\/$ are
monotonically decreasing. In the IR limit, they are finite and equal
$\alpha_{\rm an}(0,3)=\tilde{\alpha}(0,3)\simeq 1.4\,$ with the same
infinite derivatives. Both have the same leading term
$\,\sim 1/\ln x\/$ in the UV limit.\par

--- All other functions (``effective coupling powers") of both the sets
start from the zero IR values ${\cal A}_{k\geq 2}(0,f)=
{\mathfrak A}_{k\geq 2}(0,f)=0$ and obey the UV behavior
$\sim 1/(\ln x)^k$\, corresponding to $\albars^k(x)$. They are no longer
monotonous. The second functions $\,{\cal A}_2\,$ and $\,{\mathfrak A}_2\,$
are positive with maximum around $\;s, Q^2 \sim \Lambda^2$. Higher functions
$\,{\cal A}_{k\geq 3}\,$ and $\,{\mathfrak A}_{k\geq 3}\,$ oscillate in the
region of low argument values and obey $\,\/k-2\/\,\;$ zeroes. \par

  Remarkably enough, the mechanism of liberation of unphysical
singularities is quite different. While in the space-like domain
it involves non-perturbative, power in $Q^2$, structures, in the
time-like region, it is based only upon resummation of the
``$\pi^2$ terms". Figuratively, (non-perturbative !) {\it
analyticization\/} in the $Q^2$--channel can be treated as a
quantitatively distorted reflection (under $Q^2\to s=- Q^2$) of
(perturbative) {\it``pipization"} in the $s$--channel. This effect
of ``distorting mirror" first discussed in \cite{mo98} is
illustrated in figures 1 and 2. \smallskip

  Summarize the main results essential for data analysis. Instead of
power perturbative series in the space-like
$ D_{\rm pt}(Q^2)=1+d_{\rm pt}(Q^2)\:$
$$
~ d_{\rm pt}(Q^2)= \sum_{k\geq 1}^{}d_k\,\albars^k(Q^2;f)
$$
and time-like regions $ R_{\rm pt}(s)=1+r_{\rm pt}(s)\,$
$$
r_{\rm pt}(s)=\sum_{k\geq 1}r_k\, \tildal^k(s;f)\,;\quad (r_{1,2}=d_{1,2},
 r_3=d_3- d_1\frac{\pi^2\beta_{[f]}^2}{3}, \ r_4=d_4-\dots\,)\,,$$
 one has to use asymptotic expansions (\ref{Dan}) and (\ref{Rpi})
$$d_{\rm an}(Q^2)= \sum_{k\geq 1}d_k\, {\cal A}_k(Q^2,f) \,;~\quad
r_{\pi}(s)=\sum_{k\geq 1}d_k\, {\mathfrak A}_k(s,f) $$ with {\it the
same coefficients\/} $d_k$ over non-power sets of functions $\left\{{\cal
A}\right\}$ and $\left\{{\mathfrak A}\right\}$.

\section{Global formulation of APT \label{s3}}

To apply the modified APT to analyze QCD processes, it is necessary
to formulate it ``globally", for the whole domain accessible to modern
experiment, that is  for regions with various flavour numbers $\,f\,$ of
active quarks. To this goal, one has to consider the issue of heavy quark
threshold crossing.  \par \smallskip

\subsection{\sf Threshold matching. \label{ss3.1}}
 In a real calculation, the
procedure of the threshold matching is in use. One of the simplest is
the matching condition in the massless \msbar scheme\cite{match}
\begin{equation}\label{Q2match}
\albars(Q^2=M^2_f; f-1) = \albars(Q^2=M^2_f; f)\eeq
related to the mass squared $M_f^2$ of the f-th quark. \par
   This condition allows one to define a ``global" function
$\,\albars(Q^2)\,$ consisting of the smooth parts
\beglab{19}
\albars(Q^2)=\,\albars(Q^2;f)\quad\mbox{at}\quad M^2_{f-1}\leq
Q^2\leq  M^2_f \eeq
 and continuous in the whole space-like interval of positive               % c 11
$\,Q^2\,$ values with discontinuity of derivatives at the matching
points. We call such  functions  the {\it spline--continuous}~ones.

 At first sight, any massless matching, yielding the spline--type function,
violates the analyticity in the $Q^2\,$ variable, thus disturbing the
relation between the $s$-- and $Q^2\,$--channels\footnote{Any massless
scheme is an approximation that can be controlled by the related
mass--dependent scheme~\cite{dv81}. Using such a scheme, one can
devise~\cite{mikh} a smooth transition across the heavy quark threshold.
Nevertheless, from the practical point of view, it is sufficient (besides
the case of data lying in close vicinity to the threshold) to use the
spline--type matching (\ref{Q2match}) and forget about the smooth
threshold crossing.}. \par

 However, in the IAA, the original power perturbation series (\ref{Dpt})
with its unphysical singularities and possible threshold non-analyticity
has no direct relation to data, being a sort of a ``raw material" for
defining spectral density. Meanwhile, the discontinuous density is not
dangerous. Indeed, an expression of the form
\begin{equation}\label{discont}                                             %  20
\rho_k(\sigma)=\rho_k(\sigma; 3) + \sum_{f\geq 4}^{}\theta(\sigma-M_f^2)
\left\{\rho_k(\sigma; f)-\rho_k(\sigma; f-1) \right\} \end{equation}
with $~\rho_k(\sigma; f)= \Im\,\albars^k(-\sigma, f)$
 defines, according to (\ref{defAk}) and (\ref{m-s}), the smooth global
\begin{equation} \label{globalAQ}
{\cal A}_k(Q^2) =\frac{1}{\pi}\int\limits_{0}^{\infty} \frac{d\sigma}
{\sigma+x} \:\rho_k(\sigma)
\end{equation}
 and spline--continuous global
\begin{equation} \label{globalAs}
\mathfrak{A}_k(s)=\int\limits^{\infty}_{s}\frac{d \sigma}{\sigma}
\rho_k(\sigma)\, \end{equation}
 functions \footnote{Here, by eqs.(\ref{globalAQ}),(\ref{globalAs}) and
(\ref{discont}) we have introduced new ``global" effective invariant couplings
and higher expansion functions different from the previous ones with a fixed
$f$ value.}.\par

 We see that in this construction the role of the input perturbative invariant
coupling $\albars(Q^2)$ is twofold. It provides us not only with spectral
density (\ref{discont}) but with matching conditions (\ref{Q2match})
relating $\,\Lambda_f\,$ with $\,\Lambda_{f+1}\,$ as well.  \par
  Note that the matching condition (\ref{Q2match}) is tightly related
\cite{match,mikh} to the renormalization procedure. Just for this profound
reason we keep it untouched (compare with Ref.~\cite{mo98}). \smallskip

\subsection{\sf The $s$-channel: shift constants.\label{ss3.2}}

 As a practical result, we now observe that the ``global" $s$--channel
coupling $\,\tildal(s)\,$ and other functions ${\mathfrak A}_k(s)$ generally
differ from the effective coupling with a fixed flavor number $f$
$\,\tildal(s;f)\,$ and ${\mathfrak A}_k(s;f)$ by constants. For example,
at $M^2_5\leq s\leq M^2_6$
$$
\tildal(s)\,=\,\int\limits^{\infty}_{s}\frac{d \sigma}{\sigma}\rho(\sigma)\,=
\,\int\limits^{M^2_6}_{s}\frac{d \sigma}{\sigma}\rho(\sigma;5)\,+
\,\int\limits^{\infty}_{M^2_6}\frac{d\sigma}{\sigma}\rho(\sigma;6)\,=
\tildal(s; 5)\,+c(5)\,. $$
Generally,
\begin{equation}\label{alpha-s-f}
\tildal(s)\,=\tildal(s; f)\,+c(f)\,\quad \quad \mbox{at}\quad
\quad M^2_f\leq s \leq M^2_{f+1}\,  \eeq
with {\it shift constants} $c(f)$ that can be calculated in terms of
integrals over $\rho(\sigma; f+n) \,\,\, n\geq 1 \,$ with additional
reservation $\,c(6)=0\,$ related to the asymptotic freedom condition. \par
More specifically,
$$
c(f-1)=\tildal(M^2_f; f)-\tildal(M^2_f; f-1)+c(f)\,\:,\quad c(6)=0 \:. $$
  These $\,c(f)\,$ reflect the $\,\tildal(s)\,$ continuity at the
matching points $\,M^2_f$.  \par

  Analogous shift constants
\begin{equation}
{\mathfrak A}_k(s)\,=\,{\mathfrak A}_k(s; f)\,+{\mathfrak c}_k(f) \,
\quad \mbox{at} \quad M^2_f\leq s \leq M^2_{f+1}\,\end{equation}          %eq 24
are responsible for continuity of higher expansion
functions. Meanwhile, $\,{\mathfrak c}_2(f)\,$ relates to
discontinuities of the ``main" spectral function (\ref{discont}).

 The one-loop estimate with $\,\beta_{[f]}\rho(\sigma;f) =
\left\{\ln^2(\sigma/\Lambda^2_f)+ \pi^2\right\}^{-1}\,$,
\beq\label{sc}
 c(f-1)-c(f) =\frac{1}{\pi\beta_{[f]}}\arctan\frac{\pi}{\ln\frac{M^2_f}
 {\Lambda_f^2}}-\frac{1}{\pi\beta_{[f-1]}}\arctan\frac{\pi}{\ln\frac{M^2_f}
{\Lambda_{f-1}^2}} \simeq \frac{17-f}{54}\albars^3(M_f^2) ~\eeq
and traditional values of the scale parameter $\,\Lambda_{3},
\Lambda_4 \sim 350-250$ \MeV \ reveal that these constants
 $$
c(5)\simeq 3.10^{-4}\:,~c(4)\simeq 3.10^{-3}\:,~c(3)\sim
0.01\,\,;\quad \:{\mathfrak c}_2(f)\simeq 3\,\alpha(M_f^2)\,c\,(f) \,$$
are essential at a few per cent level for $\,\tildal\,$ and at ca
10\% level for $\,{\mathfrak A}_2\,$.  \par

 This means that the quantitative analysis of some $s$--channel events
like, e.g., $e^+e^-$ annihilation~\cite{ss99tmp}, $\tau$--lepton
decay~\cite{MSS-PL97-tau} and charmonium width~\cite{kras82} at the $f=3\,$
region should be influenced by these constants. \par
\smallskip

\subsection{\sf Global Euclidean functions. \label{ss3.3} }
  On the other hand, in the Euclidean, instead of the spline-type function
$\albars\,$, we have now continuous, analytic in the
whole $\,Q^2>0\,$ domain, invariant coupling defined, along with
(\ref{globalAQ}), via the spectral integral
\begin{equation} \label{an-spectr}
\alpha_{\rm an}(Q^2)=\frac{1}{\pi} \int\limits_{0}^{\infty}\frac{d \sigma}
{\sigma + Q^2}\; \rho(\sigma) \end{equation}
with the discontinuous density $\rho(\sigma)$ (\ref{discont}). \par

  Unhappily, here, unlike  the time-like region, there is no possibility
of enjoying any more explicit expression for $\,\alpha_{\rm an}(Q^2)\,$
even in the one-loop case. Moreover, the Euclidean functions
$\,\alpha_{\rm an}$ and $\,{\cal A}_k\,$, being considered in a
particular $f$--flavour region $\,M^2_f\leq Q^2 \leq M^2_{f+1}\,$, do depend
on all $\,\Lambda_3\,, \dots ,\,\Lambda_6\,$ values simultaneously. \par

 Nevertheless, the real difference from the $f=3$ case, numerically, is not
big at small $Q^2\,$ and in the ``few GeV region", for practical
reasons, it could be of importance .
%%%%%  FIGURE-2   ===  %%%%%%%%%%%%%%%%
          \begin{figure}[]\label{fig2}
 \unitlength=1mm
    \begin{picture}(0,100)                           %
    \put(10,1){%                                     %
    \epsfig{file=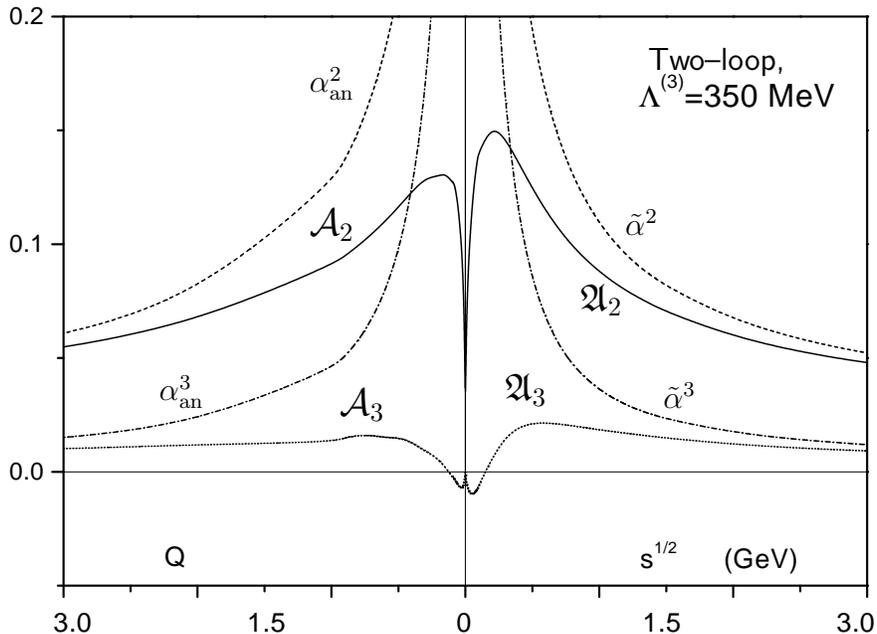,width=14.0cm}%           %
    }                                                %
   \put(108,86){\sf Two--loop,}%                     %
    \put(63,83){$\alpha_{\rm an}^2$}%                %
    \put(63,64){\large\bf ${\cal A}_2$}%             %
    \put(43,42){$\alpha_{\rm an}^3$}%                %
    \put(67,40){\bf\large ${\cal A}_3$}%             %
    \put(105,64){$\tilde{\alpha}^2$}%                %
    \put(99,54){\large ${\mathfrak A}_2$}%           %
    \put(110,41){$\tilde{\alpha}^3$}%                %
    \put(89,42){\large ${\mathfrak A}_3$}%           %
   \end{picture}                                     %
 \caption{\sl  ``Distorted mirror symmetry" for global
expansion functions}
 \end{figure}
%%%%%%%%%%%%%%%%%%%%%%%%%%%%%%%%%%

 This situation is illustrated by Fig. 2. Here, by thick solid curves with
maxima around $\sqrt{s}, Q \equiv \Lambda\,$, we draw expansion functions
${\cal A}_2$ and ${\mathfrak A}_2\,$ in a few GeV region. Thin solid lines
zeroes around $\Lambda$ and negative values below, represent ${\cal A}_3$
and ${\mathfrak A}_3\,$. For comparison, we give also second and third
powers of relevant analytic couplings \alphan and \tildal.

All these functions correspond to exact two--loop solutions expressed in terms
of Lambert function \footnote{Details of these calculations can be found in
Ref.\cite{mag00}. Assistance of D.S. Kurashev and B.A. Magradze in
calculation of curves with Lambert functions is gratefully acknowledged.}.

\section{Illustrations  \label{s4}}
  Another quantitative effect stems from the nonpower structure of
the IAA perturbative expansion. It is also emphasized at the few GeV region.

\subsection{The $s$--channel \label{ss4.1} }
 To illustrate the qualitative difference between our global scheme and
common practice of data analysis, we first consider the $f=3$ region. \medskip

 The process of \underline{\sf Inclusive $e^+e^-$ hadron annihilation}         % c 14
provides us with an important piece of information on the QCD parameters.
In the usual treatment, (see, e.g., Refs.\cite{pdg00,bard99}) the basic
relation can be presented in the form
\begin{equation}\label{Rtrad}
\frac{R(s)}{R_0} =1+ r(s)\,; \quad r_{pt}(s)=\frac{\albars(s)}{\pi}+r_2\,        %  27
\albars^2(s)+ r_3\,\albars^3(s)\,. \end{equation}
 Here, the numerical coefficients $\, r_1=1/\pi
=0.318\,,~\,r_2=0.142\,,~r_3=- 0.413\,$ (given for the $f=5$ case) are not
diminishing. However, a rather big negative $r_3$ value comes mainly from
the $\,-r_1\pi^2\beta^2_{[5]}/3$ contribution equal to $-0.456$. Instead of
(\ref{Rtrad}), with due account of (\ref{Rpi}), we now have
\begin{equation}\label{r-new}
r_{\pi}(s)= \frac{\tildal(s)}{\pi}+d_2\,{\mathfrak
A}_2(s)+d_3\:{\mathfrak A}_3(s)\:;\end{equation}
with reasonably decreasing coefficients $\,d_1=0.318\,;~d_2=0.142\,
;~d_3=0.043\,,$ the mentioned $\pi^2$ term of $r_3$ being ``swallowed" by
$\,\tildal(s)\,$\footnote{This term contributes about $8.10^{-4}$
to the $r(M^2_Z)$ and, correspondingly, 0.0025 to the extracted
$\albars(M^2_Z)$ value. This means that the main part of the
``traditional three-loop term" $r_3 \albars^3$ in the r.h.s. of
(\ref{Rtrad}), being of the one--loop origin, is essential for the
modern quantitative analysis of the data. In particular, it should
be taken into the account even in the so-called NLLA which is a
common approximation for the analysis of events at $\sqrt{s}\sim
M_Z\,.$ For a more detailed numerical APT analysis of the $f=5$
region, see \cite{pipi}.}\par
 Now, the main difference between (\ref{r-new}) and (\ref{Rtrad})
is due to the term $\,d_2\,{\mathfrak A}_2\,$ standing in the place of
$~d_2\,\tildal^2$. The difference can be estimated by adding into
(\ref{Rtrad}) the structure $\,r_4\,\alpha^4 \,$ with $\,r_4 \simeq -1.$
 This effect could be essential in the region of $\tildal(s)\simeq 0.20-0.25$.
  Here, in the APT analysis, the third, three-loop term contributes
about half of a per cent, compared with 5,5\% in the usual case.\par
\smallskip

 The APT algorithm with fixed $f=3$ has recently been used~\cite{mssy00}
for the analysis of \underline{Inclusive $\tau$--decay.} Here, the
theoretical expression for an observed quantity, the time-life of $\tau\,$
lepton, contains QCD correction $\Delta$ expressed via an integral of an
$s$--channel matrix element over the region $0< s < M_{\tau}^2$.

  As a result of the three--loop analysis of a modern \cite{exp99}
experimental value $\,\Delta_{exp}(s_0=3.16 Ē'^2)=0.191\,,$ it was obtained
that $\,\tildal(M^2_{\tau})=0.380\,.$ Remind here that under usual treatment
one  obtains $\,\albars(M^2_{\tau})=0.334\,$ that can hardly be related to
any $\,\albars(M^2_{\tau})$ value as far as the parameter
$\pi^2/\ln^2(M^2_{\tau}/\Lambda^2)$ is close to unity.\par

 Note also that the third term of (\ref{r-new}) contributes here about
one per cent. \medskip

\subsection{The $Q^2$--channel : Sum Rules \label{ss4.2}}
In the Euclidean channel, instead of power expansion like (\ref{Dpt}),
we typically have
 \begin{equation}
d(Q^2)=\frac{\alpha_{\rm an}(Q^2)}{\pi} +d_2\,{\cal A}_2(Q^2)+
 d_3\,{\cal A}_3(Q^2)\:. \end{equation}
 Here, the modification is related to a non-perturbative power
structures behaving like $\,\Lambda^2/Q^2\,$ at $\,Q^2 \gg\Lambda^2\,$.
 As it has been estimated above, these corrections could be essential
in a few \GeV \ region. \par

In the paper \cite{mss98bj}, the IAA has been applied to \underline{\sf
the Bjorken sum rules.} Here, one has to deal with the $Q^2$--channel at
small transfer momentum squared $\,Q^2 \lesssim 10\,\GeV^2\,$. \par

  Due to some controversy of experimental data, we give here only a part of
the results of \cite{mss98bj}.  For instance, using data of the SMC
Collaboration \cite{smc97} for
$Q_0^2=10\,\GeV^2\,,$ the authors obtained $\alpha_{\rm an}^{(3)}(Q_{0}^{2})=
0.301\,$ instead of $\alpha_{\rm pt}^{(3)}(Q_{0}^{2})=0.275\,$. \smallskip
Here, the contribution of the third term is also suppressed.  \par

The same remark is valid in the analysis of the Gross--Llywellin-Smith
(GLS) sum rules. Indeed, as it was shown in paper \cite{mss99gls},
instead of proportions
 $(65:24:11)_{\rm TB}$\footnote{That is, the contribution of the first, linear in
\albars, is  65 \% , while the contributions of the second and third are 24
and 11 per cent.} of usual analysis, the APT gives $(75:21:4)_{\rm APT}\,$
(for further details, see Section IIc in \cite{mss99gls}.
  The same effect for the Bjorken sum rules turns out\cite{mss98bj} to be
more pronounced $(55:26:19)_{\rm TB} \to (80:19:1)_{\rm ATB}\,.$  \medskip
\smallskip

 {\sf Some comments} are in order: \par
---  We see that, generally, the extracted values of $\alpha_{\rm an}$ and
of $\tilde{\alpha}\,$ are both slightly greater in a few GeV region than
the relevant values of $\albars$ for the same experimental input. This
corresponds to the above-mentioned non-power character of new asymptotic
expansions with a suppressed higher-loop contribution. \par

---  At the same time, for equal values of
$\,\alphan(x_*)=\tildal(x_*) =\albars(x_*)\,$, the analytic scale
parameter $\Lambda_{\rm an}\,$ values extracted from $\,\alphan\,$
and $\,\tildal\,$ are a bit greater than that $\,\Lambda_{\msbar}\,$ taken
from $\albars$. This feature is related to a ``smoother" behavior of
both the regular functions $\alphan$ and $\tildal\,,$ as compared
to the singular $\albars$. \par \smallskip

\subsection{Conclusion \label{ss4.3}}
 To summarize, we repeat once more our main points.\smallskip

 1. We have formulated a self-consistent scheme for analyzing data in both
the space-like and time-like regions. \smallskip

 The fundamental equation connecting these regions is the dipole
spectral relation (\ref{d-trans}) between renormalization--group
invariant nonpower expansions $D_{\rm an}(Q^2)$ and $R_{\pi}(s)$.\par

 Just this equation (equivalent to the K\"allen--Lehmann representation),
treated as a transformation,  is responsible for non-perturbative terms
in the $\/Q^2\,$--channel involved into $\alphan(Q^2)$ and non-power
expansion functions $\left\{{\cal A}_k(Q^2)\right\}$. These terms,
non-analytic in the coupling constant $\alpha$, are a counterpart
to the perfectly perturbative $\,\pi^2$--terms effectively summed
in the $s$--channel expressions \tildals and $\left\{{\mathfrak
A}_k(s)\right\}$.\smallskip

2. As a by-product, we ascertain a new qualitative feature of the
IAA, relating to its non-perturbativity in the $Q^2$--domain. It
can be considered as a {\sf minimal non-perturbativity} or minimal
non-analyticity\footnote{Compatible with the RG invariance and the
$Q^2$ analyticity --- compare with \cite{dv76}.} in $\alpha$ as
far as it corresponds to perturbativity in the $s$--channel. \par

 Physically, it implies that minimal non-perturbativity cannot be
referred to any mechanism producing effect in the
$s$--channel.\smallskip

 3. The next result relates o the correlation between regions with
different values of the effective flavor number $f\,$. Dealing
with the massless \msbar renormalization scheme, we argue that the
usual perturbative QCD expansion provides our scheme only  with
step--discontinuous spectral density (\ref{discont}) depending
simultaneously on different scale parameters $\,\Lambda_f\,;~f=
3,\dots,6\,$ connected by usual matching relations.\par

  This step--discontinuous spectral density yields, on the one hand, %{\sf
smooth analytic coupling $\alpha_{\rm an}(Q^2)\,$ and higher functions
$\left\{{\cal A}_k(Q^2)\right\}\,$ in the space-like region%}
 --- eq.(\ref{globalAQ}).\par
 On the other hand, it produces the  spline--continuous invariant
coupling $\,\tilde{\alpha}(s)\,$  and functions $\left\{{\mathfrak
A}_k(s)\right\}$ in the time-like region --- eq.(\ref{globalAs}).

  As a result, the global expansion functions $\left\{{\cal A}_k(Q^2)
\right\}\,$ and $\left\{{\mathfrak A}_k(s)\right\}\,$ differ both from
the  ones $\left\{{\cal A}_k(Q^2;f)\right\}\,$ and
$\left\{{\mathfrak A}_k(s;f)\right\}\,$ with a fixed value of a
flavour number.\smallskip

4. Thus, our global APT scheme uses the common invariant coupling
$\,\albars(Q^2, f)\,$ and matching relations, only as an input. Practical
calculation for an observable now involves expansions over the sets
$\,\left\{{\cal A}_k(Q^2)\right\}\,$ and $\,\left\{\agothk(s)\right\}\,,$
that is non-power series with usual numerical coefficients $\,d_k\,$
obtained by calculation of the relevant Feynman diagrams. \par

   In particular, this means that we have now {\sf three QCD effective
couplings:} \tildal, \alphan --- of the APT formalism, and traditional
$\albars,.$ This usual one can be used for approximate expression of two
first ones in four and five--flavor regions, for the comparison reasons. \par

  This means that, generally, one should check the accuracy of the bulk
of extractions of the QCD parameters from diverse ``low energy" experiments.
Our preliminary estimate shows that such a revision could influence the rate
of their correlation. \smallskip

5. Last but not least. As it has been mentioned in our recent publications
~\cite{prl97,ss99tmp}, the IAA obeys  immunity with respect to higher
loop and renormalization scheme effects. \par
  Now, we have got an additional insight into this item related to observables
and can state that the perturbation series for an observable in the IAA have
better convergence properties (than in the usual RG--summed perturbation theory)
in both the $s$-- and $Q^2\,$-- channels. \medskip

{\bf\large Acknowledgements} \smallskip

The author is indebted to D.Yu.~Bardin, N.V.~Krasnikov, D.S. Kurachev,
B.A.~Magradze, S.V. Mi\-k\-hai\-lov, A.V. Radyushkin, I.L. Solovtsov and
O.P. Solovtsova
for useful discussion and comments. This work was partially supported by
grants of the Russian Foundation for Basic Research (RFBR projects Nos
99-01-00091 and 00-15-96691), by INTAS grant No 96-0842 and by INTAS-CERN
grant No 2000-377. %)newpage
\tableofcontents
\end{document}